\documentclass[letter]{my_aa}
\usepackage{graphicx,natbib,txfonts}
\bibpunct{(}{)}{;}{a}{}{,} 

\begin{document}

\title{Probing the dusty environment of the Seyfert~1 nucleus in NGC~3783 with MIDI/VLTI interferometry}
\author{
        Thomas~Beckert
           \and 
        Thomas~Driebe
           \and 
        Sebastian~F.~H{\"o}nig
           \and
        Gerd~Weigelt
          }
\institute{Max-Planck-Institut f\"ur Radioastronomie, Auf dem H\"ugel 69,
           53121 Bonn, Germany
          }
\date{}

\abstract {}
{We present mid-IR spectro-interferometry of the Seyfert type 1 nucleus of \object{NGC 3783}. The dusty circumnuclear environment is spatially resolved and the wavelength dependence of the compact emission is discussed.} 
{The observations were carried out with the MIDI instrument at the Very Large Telescope Interferometer in the $N$-band. 
Spectra and visibilities were derived with a spectral resolution of $\lambda/\Delta\lambda\sim30$ in the wavelength range from 8 to $13\,\mu$m. For the interpretation we developed a simple dusty disk model with small and variable covering factor.}
{At baselines of 65 and 69\,m, visibilities in the range of 0.4 to 0.7 were measured. The $N$-band spectra show a monotonic increase of the measured flux with wavelength with no apparent silicate feature around $10\,\mu$m. We find that the mid-IR emission from the nucleus 
can be reproduced by an extended dust disk or torus with a small covering factor of the radiating dust clouds.
} {Our mid-IR observations of NGC~3783 are consistent with a clumpy circumnuclear dust environment. {  The interpretation in terms of a dusty torus with low covering factor supports a clumpy version of the unified scheme for AGN}. The inferred sizes and luminosities are in good agreement with dust reverberation sizes and bolometric luminosities from optical and X-ray observations. }
\maketitle

\section{Introduction}
Unified scenarios of AGN suggest that the central continuum source
is surrounded by a dusty molecular torus.
This torus is thought to be responsible for aspect-angle-dependent
obscuration of the central source \citep[e.g.,][]{Antonucci1993ARAA_31_473}.
Mid-infrared (MIR) interferometry of the Seyfert~2 nucleus in \object{NGC 1068} 
\citep{Jaffe2004,Poncelet2006AA.450.483} and the \object{Circinus galaxy} \citep{Tristram2007AA.474.837} provided supporting evidence for the dusty torus model. The MIR emission of these nuclei was resolved by MIR interferometry with the MIDI instrument in both cases. The resolved structures are belived to be the IR-bright innermost region of the torus. In Seyfert type 1 sources the situation is not as clear as NIR interferometry data obtained with the Keck interferometer on \object{NGC 4151} \citep{Swain2003ApJ.596L.163} have been interpreted as accretion disk emission.

At a redshift of $z=0.0097$ \citep[][$D = 44$\,Mpc]{Theureau1998AAS_130_333}, the galaxy NGC~3783 
harbors a bright Seyfert 1 core 
with a black hole of $M_{\rm BH} = 3.0 \pm 0.5 \times 10^7 M_\odot$ derived from 
reverberation mapping \citep{Onken2004ApJ_615_645}.
NGC~3783 has been the target of extensive ultraviolet spectral monitoring \citep[e.g.,][]{Reichert1994ApJ_425_582}.
A narrow iron K line is seen in XMM-Newton spectra \citep{Reeves2004ApJ_602_648} consistent with an origin in Compton-thick matter such as a molecular and dusty torus.
The optical spectrum shows blueward asymmetries in forbidden lines and broad lines with up to 3800 km/s in the broadest components \citep{Evans1988ApJS_67_373}. 
The  extended narrow line region of NGC~3783 seen with HST \citep{Schmitt2003ApJS_148_327} appears halo-like in [O{\sc iii}] with a size of $1\farcs9$\, (400 pc).
The NIR spectrum of the nucleus \citep{Reunanen2003MNRAS_343_192} features a broad  Br$\gamma$ line and three coronal lines are detected.
MIR spectra within a beam of $4\farcs2$\, were presented by \citet{Roche1991MNRAS_248_606} and showed an almost featureless spectrum with a $N$-band luminosity of $\sim 3\times 10^9 L_\odot$ in this aperture. 
VISIR photometry presented by \citet{Horst2006AA_457L_17} show that NGC~3783 falls onto the MIR vs. hard X-ray correlation for AGN with a rising flux of 0.66 to 0.72\,Jy across the $N$-band.
The UV luminosity is estimated from $L_{\rm UV} \approx 6\,\lambda L_\lambda (5100$\,\AA) to be $1-2\times 10^{37}\,$W. This is consitent with the $L_{\rm bol} - L_{\rm X}$-relation \citep{Marconi2004MNRAS_351_169} which predicts a bolometric luminosity of $3 \times 10^{37}$\,W for NGC~3783. The accretion disk around the central black hole is therefore radiating at $\sim 10\%$ of the Eddington limit. 
All these characteristics of a classical type 1 core of suggest the presence of a dusty nuclear torus seen almost face-on. Direct evidence of this torus can only be obtained through high spatial resolution interferometry of the hot dust emission in the near- and mid-IR.  
In this paper we present the first spectro-interferometric observations of NGC~3783 with MIDI/VLTI. In Sec.\,2 we describe the observations and data reduction. In Sec.\,3 we develop a simple model for the circumnuclear dust distribution and discuss the implications for dusty tori in the unification scheme of AGN. Sec.\,4 gives a short summary.

\section{Observations}
NGC~3783 was observed with MIDI in the two nights 27 and 30 of May 2005 (Program 75.B-0697). A prism with a spectral resolution of $\lambda/\Delta \lambda = 30$ at $10\,\mu$m was used to obtain spectrally dispersed fringes between 8 and $13\,\mu$m. A detailed description of the observing procedure is given in \citet{Przygodda2003ApSS_286_85}. 
We use the data reduction package EWS developed at Leiden Observatory (combined MIA+EWS, ver.1.5.2)\footnote{Available at http://www.mpia-hd.mpg.de/MIDISOFT/ and http://www.strw.leidenuniv.nl/nevec/MIDI/index.html.}. The EWS software first corrects for instrumental and atmospheric path differences before adding the fringes coherently \citep{Jaffe2004SPIE}. We derive the interferometer transfer function at each wavelength (i.e., at each spectral channel) between 8 and $13\,\mu$m by observing calibrators whose angular diameters are known (see Table\,\ref{table:0}). 

\begin{table}
\caption{List of calibrators used in this work, together with 12 $\mu $m fluxes ($F_{12}$), and uniform-disk diameters ( $d_{\rm UD}$) of the MIDI observations.} 
\label{table:0} 
\begin{center} 
\begin{tabular}{c c c c c} 
\hline\hline 
Calibrator & $F_{12}$ (Jy) &	$d_{\rm UD}$ (mas) \\ \hline
HD\,100407 & 11.6 & 2.41 \\
HD\,116870 & 10.0 & 2.57 \\
HD\,123139 & 54.4 & 5.25 \\
HD\,169916 & 22.4 & 3.90 \\
HD\,160668 & 28.8 & 2.30 \\
HD\,152885 & 12.6 & 2.88 \\
HD\,178345 & 10.6 & 2.50 \\
\hline 
\end{tabular}
\end{center}
\end{table}
In order to calibrate the visibility of NGC~3783 at a given wavelength, we use the mean of the transfer function values derived from all the calibrators observed on the same night as NGC~3783. The errors of the calibrated visibilities are estimated from errors on the raw visibilities and the standard deviation of the transfer function values at each wavelength (added in quadrature). 
The relative errors of the calibrated visibilities are typically $\sim 10\%$. 
\begin{figure}
 \resizebox{\hsize}{!}{\includegraphics[angle=0]{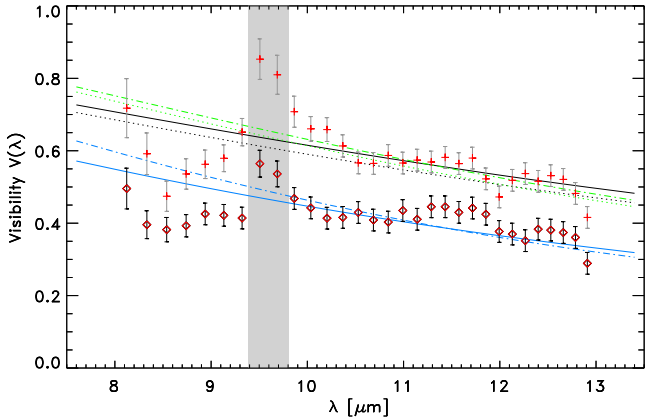}}
 \caption{\label{N3783VisC} Visibilities vs. wavelength of NGC~3783 in the $8-13\,\mu$m range observed with MIDI on 30 May 2005. Visibilities from dataset 1 (diamonds) show smaller visibilities than dataset 2 (crosses) for projected baselines of $B=68.6$\,m at P.A.$=115\degr$ and  $B=64.9$\,m at P.A.$=120\degr$, respectively. The gray area between $\lambda = 9.4$ and $9.8\,\mu$m is affected by the atmospheric ozone band, which could not be properly calibrated. The model visibilities A-1 (blue, dash-dotted) A-2 (green, dash-dotted), B-1 (blue, solid) and B-2( black,solid) are shown for comparision. The models A-2 (green) and B-2 (black), modified for the longer baseline of $B=68.6$\,m, are also plotted as dotted curves. This demostrates that the visibilty offset between No.~1 and 2 is not explained by the changing baseline when the visibility fo each dataset utilizes the associated photometry shown in Fig.~\ref{N3783Spec}.}
\end{figure}
Fig.~\ref{N3783VisC} shows the calibrated visibilities of NGC~3783 from two datasets (No.\,1\&\,2 in the following) obtained close in time in the night 30 May 2005. The observed $N$-band visibilities are in the range of 0.4 to 0.7. The wavelength dependence is weak within the errors from 8 to $13\,\mu$m for both datasets. The projected baselines are $B = 68.6$\,m and $64.9\,$m for the dataset 1 and 2, respectively. We include in Fig.~\ref{N3783VisC} also the visibilities of two different models described in Sec.\,\ref{Model:Sec} below. The wavelength region around $9.6\,\mu$m (marked as a gray area in Figs.\,\ref{N3783VisC} and \ref{N3783SpecComp}) is difficult to calibrate for faint targets with $F_\nu \approx 1\,$Jy due to a strong atmospheric ozone band. This may explain the apparently larger visibilities in that wavelength region.
\begin{figure}
 \resizebox{\hsize}{!}{\includegraphics[angle=0]{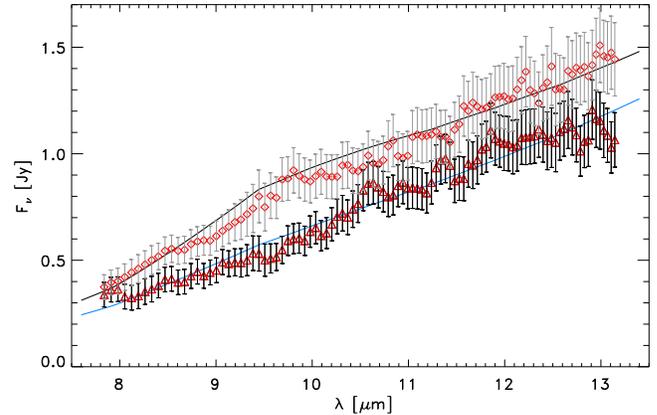}}
 \caption{\label{N3783Spec} MIDI $N$-band spectra of the nucleus of NGC~3783 in Jy vs. wavelength. The spectra have been calibrated and averaged against the reference stars of the night 30 May 2005. Dataset 1 and 2 are shown with diamonds (1) and triangles (2), respectively. The model spectra for B-1 (blue) and B-2 (black) of a clumpy and dusty torus corresponding to Fig.\,1 and described in Sec.\,\ref{Model:Sec} are overplotted. The weak or absent silicate feature around $\lambda=9.7\,\mu$m is a consequence of the large optical depth in the models.}
\end{figure}
In addition, we extracted absolute $N$-band spectra of NGC~3783 from MIDI single channel photometry
Interferometric calibrators observed in the same night as NGC~3783 and at similar airmasses serve as spectrophotometric standard stars to calibrate the spectra (see Table\,\ref{table:0}). The calibrated spectra of these stars are taken from \citet{Cohen1999AJ_117_1864}. 
The two calibrated spectra of NGC~3783 are shown in Fig.~\ref{N3783Spec}. We estimate the errors of the spectra to be 10-20\%.
The flux spectrum rises slowly with wavelength over the whole $8-13\,\mu$m range. It is important to note that neither the flux, in agreement with \citet{Roche1991MNRAS_248_606}, nor the visibility spectrum shows a prominent silicate feature at $\sim 10\,\mu$m. 

{  Due to variable thermal emission from the sky and from inside the interferometer, the uncorrelated photometry measurements are subject to larger errors than the correlated flux measurement. In an alternative approach we combined the spectra from both datasets to create a common total flux spectrum shown in Fig.~\ref{N3783SpecComp}. We used this to derive corrected visibilities for datasets 1 and 2 also included in Fig.~\ref{N3783SpecComp}. A comparision with model B-3 shows that the remaining difference between the two datasets can be attributed to the difference in the baseline. }
\begin{figure}
 \resizebox{\hsize}{!}{\includegraphics[angle=0]{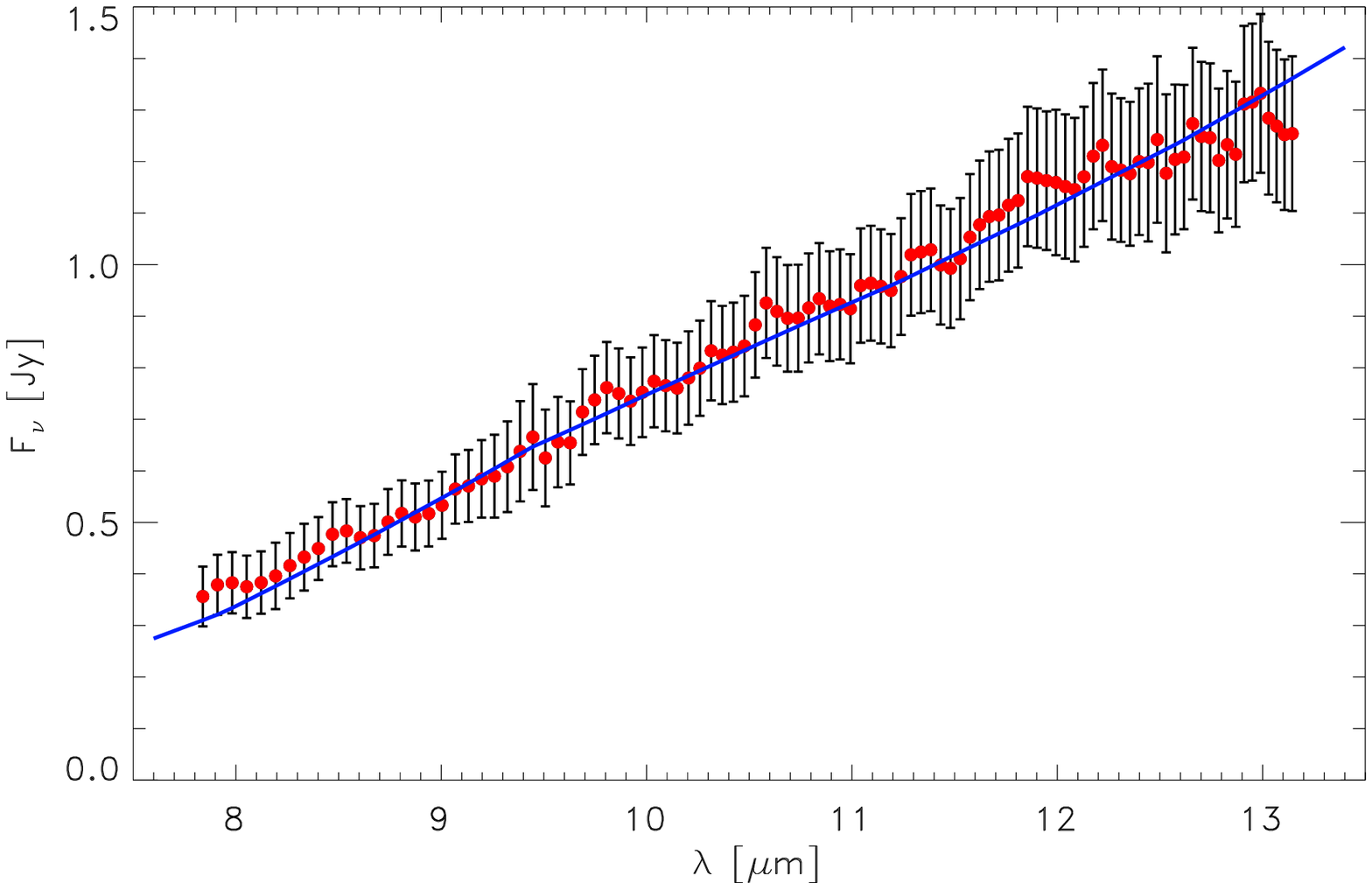}}
 \resizebox{\hsize}{!}{\includegraphics[angle=0]{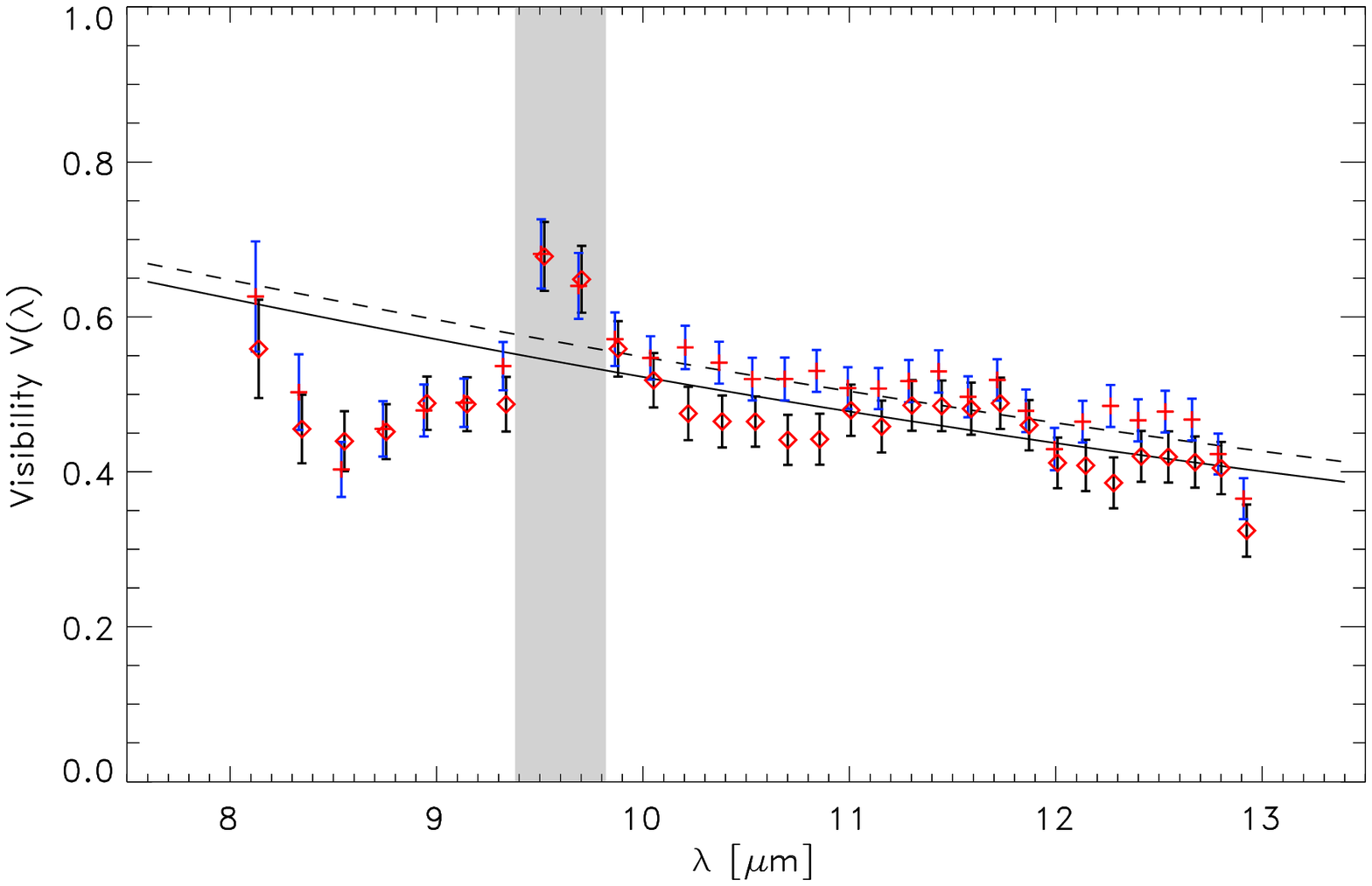}}
 \caption{\label{N3783SpecComp} Combined MIDI $N$-band spectrum (top) and visibilities (bottom) of NGC~3783. We used an arithmetic mean with errors for the two spectra from Fig.~\ref{N3783Spec} and compare it with model B-3 (blue solid curve). The visibilities for datasets 1 (diamonds) and 2 (crosses) have been derived with the combined spectrum shown at the top. The visibilities for model B-3 for baseline $B = 68.6$\,m (solid) and $64.9\,$m (dashed) are included. The measurements deviate substantially from the model in the 8.3 to $9.3\,\mu$m region.}
\end{figure}

\section{Discussion}
\subsection{General properties}
The first and most notable feature of the visibilities in Fig.\,\ref{N3783VisC} and the spectra in Fig.\,\ref{N3783Spec} is the absence of the $9.7\,\mu$m absorption or emission feature of silicate dust expected to be present in the torus. This absence may be due to missing silicates in AGN dust, because of the lower sublimation temperature $T_{\rm Si} \sim 1000\,$K when compared with graphite grains $T_{\rm C} \sim 1500\,$K. But the mid-IR emission is expected to come from $T\sim 300\ldots 800\,$K dust, where both kinds of dust grains should be present. Alternatively, the dust grain size distribution maybe dominated by large grains, for which the silicate feature is less prominent. We have no direct evidence for this in NGC~3783, except for dust-reverberation arguments (see below). The third possibility, and our choice for the models below in Sec.\,\ref{Model:Sec}, is a radiative transfer effect: For moderate optical depth in the 8--$13\,\mu$m range with $\tau \sim 0.5$ and small temperature variations along the line of sight through the dusty torus, the emission spectrum approaches a black body and the feature is suppressed. 

Secondly, the flux spectrum $F_\nu$ in Figs.\,\ref{N3783Spec} and \ref{N3783SpecComp} shows a linear rise with wavelength, in contrast to most radiative transfer simulation of face-on dusty AGN tori \citep[e.g.,]{Honig2006AA_452_459}. This may indicate the presence of additional, spatially distributed heating sources, which can increase the amount of dust at intermediate temperatures of 150 to 300\,K. Another possibility is a smaller covering factor of hot dust compared to cooler dust in a disk heated by the central engine. This idea is employed in the models below.

In addition, we find almost constant visibilities between 8--$13\,\mu$m suggesting a linear rise of the source size with wavelength. The derived visibilities are smaller than expected for a compact AGN dust torus, for which the NIR and MIR emission is dominated by a small region close to the dust sublimation radius. The diameter of the inner rim of the torus is expected (see Eq.\,\ref{Temp:Eq} below) to be $2.1\,$mas corresponding to $V$ close to unity. A small dust covering factor in the models below can explain the data and imply a large extension of the MIR emission region.

Information on the torus inclination can be infered from spectropolarimetry in the optical 400 to 500\,nm region, which shows a relatively large polarisation fraction of $\sim 1$\% 
at P.A.$\,\sim 130\degr$ \citep[][and Kishimoto priv.comm.]{Lira2007ASPC..373..407L}. The axis of the optical scattering region is therefore inclined with respect to our line of sight suggesting the minor axis of an elongated MIR disk to be close to the P.A.$\sim120\degr$ of our observations. The P.A. of the interferometric baselines for the datasets 1\,\&\,2 are very similar, so that we cannot constrain any non-circularity of the MIR emission from our data.

\subsection{Modeling visibilities and spectra with a dusty disk \label{Model:Sec}}
For the interpretation of the data presented above we employ a simplified version of the clumpy torus model \citep{Beckert2004AA_426_445,Honig2006AA_452_459}. A torus seen almost face-on ($i \approx 0\degr$) appears as a disk composed of dusty clouds. The inclination has almost no observable effect on spectra and colors as long as $i$ is smaller than the half opening angle of the torus. We assume single temperature clouds, where the cloud temperature depends only on the radius in the disk. 
For simplicity and comparison with other models we use a MRN \citep{Mathis1977ApJ_217_425} grain size distribution between 
$0.005\,\mu$m and $0.25\,\mu$m.
The inner radius of the disk or torus is at $R_{\rm in}$, where we assume a dust temperature of $T_{\rm in} = 1500\,$K, which is suggested by NIR colors of type 1 AGN \citep{Kishimoto2007AA.476.713}. The local intensity $I_\nu(r) $ emitted by clouds is a modified black body $B(\nu,T)$ at temperature $T = T_{\rm in} \times (r/R_{\rm in})^{-\eta}$ with optical depth $\tau_\nu$, so that
\begin{equation}
  I_\nu(r) = B(\nu,T(r))\left(1-e^{-\tau_\nu}\right)\;.
\end{equation}
We are interested in lines of sight through the disk parallel to the disk symmetry axis. For simplicity, we use a constant $\tau_\nu$ independent of the radius. The frequency dependence of $\tau_\nu$ is derived from the MRN distribution and a mixture of 50\% silicates and 50\% graphite grains with the optical properties taken from \citet[][and references therein]{Weingartner2001ApJ_548_296}\footnote{Files with tabulated optical constants have been taken from http://www.astro.princeton.edu/$\sim$draine/dust/dust.diel.html.}. For comparison we fix $\tau_\nu$ in the optical $\tau_V$ at $\lambda=560\,$nm. For normal black body emission in radiative equilibrium with the central illuminating UV/optical radiation we expect an exponent $\eta = 0.5$. This leads to an almost constant flux spectrum $F_\nu$ when integrated over the disk surface.
We employ two modifications (models A and B) to fit the total flux spectrum in Fig.~\ref{N3783Spec} (dash-dotted lines). Model A-1 \& A-2 for dataset 1 \& 2 assumes a different exponent $\eta =0.35$ (see Table\,\ref{table:1}). This is motivated by the
equilibrium temperature of graphite grains \citep[see e.g.,][]{Barvainis1987ApJ_320_537,Kishimoto2007AA.476.713}
\begin{equation}\label{Temp:Eq}
  \frac{T}{1500\,{\rm K}} = \left[\frac{L}{10^{37}\,{\rm W}}\right]^{0.179} \left[\frac{R}{0.13\,{\rm pc}}\right]^{-0.357}\left[\frac{\tilde{a}}{0.05\,\mu{\rm m}}\right]^{-0.179}
\end{equation}
where the grain size  $\tilde{a} = \left<a^3\right>/\left<a^2\right>$ is an average size for the grain size distribution.
Such a model can explain the spectra in Fig.\ref{N3783Spec} but not the visibilities as shown in Fig.\,\ref{N3783VisC}. We can accommodate the suggested large size of the dust emitting region by introducing a covering factor $f(r)$, for which we assume a radial dependence
$f(r) = f_0 (r/R_{\rm in})^\xi$. For models A, we use $\xi =0$ and $f_0$ can be adjusted to get the size of the source. The idea of a low covering factor is a direct consequence of clumpy torus models \citep[see e.g.,][]{Honig2006AA_452_459}. In this scenario only clouds directly illuminated by the central source contribute significantly to the torus emission and due to the clumpy structure there is a non-zero probability for such clouds at all radii.

The implied AGN luminosity for models A-1 \& A-2 are much smaller (see Tabel \ref{table:1}) than derived from optical and X-ray observations. We therefore concentrate in the following on 
models B-1 \& B-2 which provide a better fit to the spectra in Fig.\ref{N3783Spec} and the visibilities in Fig.\,\ref{N3783VisC} with a running covering factor.
All models exhibit decreasing visibilities with wavelength in Fig.\,\ref{N3783VisC} and
both models A and B give acceptable fits to the datasets 1 \& 2. The parameters of the models are summarized in Table\,\ref{table:1}.
The inner torus radii of the models are determined by the total flux level and the covering factor in the combination
$R_{\rm in}/\sqrt{f_0}$. The AGN luminosity (assuming isotropic emission) in Table\,\ref{table:1} is derived from Eq.\,\ref{Temp:Eq}. The larger luminosities for B-1, B-2 and B-3 are a direct consequence of the smaller covering factor at the inner boundary $R_{\rm in}$. Note that the angular scale for $R_{\rm in}=0.1\,$pc 
is only 0.47\,mas. The actual $T=300\,$K diameter is as large as $70\,$mas for model B-1 or 60\,mas for A-1. 

\begin{table}
	\caption{Dusty disk/torus model results. The covering factor $f_0$ and the exponents $\xi$  and $\eta$ for the covering factor and dust temperature are described in Sec.\,\ref{Model:Sec}. $\tau_V$ is the optical depth along a vertical path through clouds in the disk taken at $560\,$nm. 
} 
\label{table:1} 
\begin{center} 
\begin{tabular}{c c c c c c c} 
\hline\hline 
Model & $\tau_V$ & $f_0$ & $\xi$ & $\eta$ & $R_{\rm in}$ & $L_{\rm bol}$ \\ 
& & & & & [pc] & [$10^{37}$\,W] \\
\hline 
A-1 & 6 & 0.04 & 0 &$0.35$ & 0.064 & 0.16 \\
A-2 & 12 & 0.06 & 0 & $0.35$ & 0.04 & 0.065 \\
B-1 & 4 & $4.0\,10^{-3}$ & 0.8  & $0.5$ & 0.30 & 3.7 \\ 
B-2 & 8 & $3.8\,10^{-3}$  & 0.9 & $0.5$ & 0.21 & 1.7 \\
B-3 & 8 & $3.2\,10^{-3}$  & 0.9 & $0.5$ & 0.23 & 2.2 \\
\hline 
\end{tabular}
\end{center}
\end{table}
{  The visibilities derived with the combined total flux spectrum in Fig.~\ref{N3783SpecComp} are much more consistent with each other. The difference between dataset 1 \& 2 can be completely attributed to the difference in baseline length. This allows to fit the long wavelength side the NGC~3783 datasets with a single model B-3 which is very similiar to B-2 (see Table~\ref{table:1}). In the 8.3 to $9.3\,\mu$m region the derived $V(\lambda)$ is substantially lower than suggested by all simple power-law models. NGC~3783 appears to be even larger than the models in this wavelength region.}

\subsection{Implications for unified models}
Models of dusty nuclear tori suggested by the unified scheme 
of AGN \citep{Antonucci1993ARAA_31_473} require the dust 
to be contained in 
individual clouds or clumps \citep{Krolik1988ApJ_329_702}. 
In recent years, models for the cloud size and cloud distribution \citep{Beckert2004AA_426_445} and for the emission from a clumpy dust 
distribution \citep{Nenkova2002ApJ_570L_9,Honig2006AA_452_459} have been developed. The dynamical model based on cloud collisions \citep{Beckert2004AA_426_445} suggests that a geometrically thick and obscuring torus is expected for high mass accretion rates close to the Eddington limit of the central black hole. From the X-ray and optical/UV luminosity of NGC~3783 an Eddington ratio of 10\% has been inferred. Our simple model for the mid-IR emission suggests that the 'torus' has a much smaller covering factor than expected for the collisional scenario.

Long term optical/IR photometry of NGC~3783   
\citep[][and references therein]{Glass2004MNRAS_350_1049} suggests a time delay of 190\,d between $U$- and $L$-band\footnote{In an earlier paper \citep{Glass1992MNRAS_256P_23} a time delay of 85\,d was suggested, which would imply a much smaller physical scale for the response.}. This implies a physical path difference of $0.16\,$pc or an angular scale of 0.75\,mas, if the response is in the plane of the torus/disk. Given the uncertainties 
entering in Eq.\,\ref{Temp:Eq} we get an agreement with the B-models within a factor of 2. {  For comparision, the NIR interferometry of NGC~4151 by \citet{Swain2003ApJ.596L.163} suggests a more compact emission region like an accretion disk or a ring at $R_{\rm in}$. }
A discussion of dust sublimation radii, dust reverberation time delay and luminosity dependence can be found in \citet{Kishimoto2007AA.476.713}. The inferred AGN luminosity from the B-models (Table \ref{table:1}) of $L_{\rm bol} \approx 3\,10^{37}\,$W or from the reverberation size via Eq.\,\ref{Temp:Eq} are in nice agreement with the $L_{\rm bol} - L_{\rm X}$-relation \citep{Marconi2004MNRAS_351_169}.

\section{Summary}
We present MIDI/VLTI spectro-interferometry of the Seyfert~1 nucleus of NGC~3783 and spatially resolve the object between 8 and 13 $\mu $m. The observed visibilities suggest a significantly larger MIR emission region than the estimated dust sublimation size of 1.4\,mas. Spectra and visibilities are noticeable for the apparent absence of a silicate feature around $10\,\mu$m. In addition the flux spectrum shows a linear increase with wavelength throughout the $N$-band.
{  Consistent results are obtained with the use of an average total flux spectrum (Fig.~\ref{N3783SpecComp}) which exemplifies that the incoherent flux measurement contributs the largest uncertainties.}
Our simple modeling reproduces simultaneously the observed visibilities and spectra by a dusty disk with small covering factor seen face-on with a moderate optical depth. This model is in line with recent clumpy torus models, albeit the covering factor of clouds in NGC~3783, which contribute to the MIR emission, is very small and rises outwards in the disk or torus. 

Future near-IR and additional mid-IR interferometric measurements with different baseline length and covering a wider range of position angle are needed to confirm the present results and probe the validity of the simple models presented here. 

\begin{acknowledgements} T.B. likes to thank Christian Hummel and Sebastien Morel at ESO/Paranal for help and advice during the observations. We like to thank Makoto Kishimoto for stimulating discussions and helpful comments.
\end{acknowledgements}

\bibliographystyle{aa}
\bibliography{N3783}

\begin{thebibliography}{29}
\expandafter\ifx\csname natexlab\endcsname\relax\def\natexlab#1{#1}\fi

\bibitem[{{Antonucci}(1993)}]{Antonucci1993ARAA_31_473}
{Antonucci}, R. 1993, \araa, 31, 473

\bibitem[{{Barvainis}(1987)}]{Barvainis1987ApJ_320_537}
{Barvainis}, R. 1987, \apj, 320, 537

\bibitem[{{Beckert} \& {Duschl}(2004)}]{Beckert2004AA_426_445}
{Beckert}, T. \& {Duschl}, W.~J. 2004, \aap, 426, 445

\bibitem[{{Cohen} {et~al.}(1999){Cohen}, {Walker}, {Carter}, {Hammersley},
  {Kidger}, \& {Noguchi}}]{Cohen1999AJ_117_1864}
{Cohen}, M., {Walker}, R.~G., {Carter}, B., {et~al.} 1999, \aj, 117, 1864

\bibitem[{{Evans}(1988)}]{Evans1988ApJS_67_373}
{Evans}, I.~N. 1988, \apjs, 67, 373

\bibitem[{{Glass}(1992)}]{Glass1992MNRAS_256P_23}
{Glass}, I.~S. 1992, \mnras, 256, 23P

\bibitem[{{Glass}(2004)}]{Glass2004MNRAS_350_1049}
{Glass}, I.~S. 2004, \mnras, 350, 1049

\bibitem[{{H{\"o}nig} {et~al.}(2006){H{\"o}nig}, {Beckert}, {Ohnaka}, \&
  {Weigelt}}]{Honig2006AA_452_459}
{H{\"o}nig}, S.~F., {Beckert}, T., {Ohnaka}, K., \& {Weigelt}, G. 2006, \aap,
  452, 459

\bibitem[{{Horst} {et~al.}(2006){Horst}, {Smette}, {Gandhi}, \&
  {Duschl}}]{Horst2006AA_457L_17}
{Horst}, H., {Smette}, A., {Gandhi}, P., \& {Duschl}, W.~J. 2006, \aap, 457,
  L17

\bibitem[{{Jaffe}(2004)}]{Jaffe2004SPIE}
{Jaffe}, W. 2004, SPIE Proc., 5491, 715

\bibitem[{{Jaffe} {et~al.}(2004){Jaffe}, {Meisenheimer}, {R{\"o}ttgering},
  {Leinert}, {Richichi}, {Chesneau}, {Fraix-Burnet}, {Glazenborg-Kluttig},
  {Granato}, {Graser}, {Heijligers}, {K{\"o}hler}, {Malbet}, {Miley},
  {Paresce}, {Pel}, {Perrin}, {Przygodda}, {Schoeller}, {Sol}, {Waters},
  {Weigelt}, {Woillez}, \& {de Zeeuw}}]{Jaffe2004}
{Jaffe}, W., {Meisenheimer}, K., {R{\"o}ttgering}, H.~J.~A., {et~al.} 2004,
  \nat, 429, 47

\bibitem[{{Kishimoto} {et~al.}(2007){Kishimoto}, {H{\"o}nig}, {Beckert}, \&
  {Weigelt}}]{Kishimoto2007AA.476.713}
{Kishimoto}, M., {H{\"o}nig}, S.~F., {Beckert}, T., \& {Weigelt}, G. 2007,
  \aap, 476, 713

\bibitem[{{Krolik} \& {Begelman}(1988)}]{Krolik1988ApJ_329_702}
{Krolik}, J.~H. \& {Begelman}, M.~C. 1988, \apj, 329, 702

\bibitem[{{Lira} {et~al.}(2007){Lira}, {Kishimoto}, {Robinson}, {Young},
  {Axon}, {Elvis}, {Lawrence}, \& {Peterson}}]{Lira2007ASPC..373..407L}
{Lira}, P., {Kishimoto}, M., {Robinson}, A., {et~al.} 2007, in Astronomical
  Society of the Pacific Conference Series, Vol. 373, The Central Engine of
  Active Galactic Nuclei, ed. L.~C. {Ho} \& J.-W. {Wang}, 407--+

\bibitem[{{Marconi} {et~al.}(2004){Marconi}, {Risaliti}, {Gilli}, {Hunt},
  {Maiolino}, \& {Salvati}}]{Marconi2004MNRAS_351_169}
{Marconi}, A., {Risaliti}, G., {Gilli}, R., {et~al.} 2004, \mnras, 351, 169

\bibitem[{{Mathis} {et~al.}(1977){Mathis}, {Rumpl}, \&
  {Nordsieck}}]{Mathis1977ApJ_217_425}
{Mathis}, J.~S., {Rumpl}, W., \& {Nordsieck}, K.~H. 1977, \apj, 217, 425

\bibitem[{{Nenkova} {et~al.}(2002){Nenkova}, {Ivezi{\'c}}, \&
  {Elitzur}}]{Nenkova2002ApJ_570L_9}
{Nenkova}, M., {Ivezi{\'c}}, {\v Z}., \& {Elitzur}, M. 2002, \apjl, 570, L9

\bibitem[{{Onken} {et~al.}(2004){Onken}, {Ferrarese}, {Merritt}, {Peterson},
  {Pogge}, {Vestergaard}, \& {Wandel}}]{Onken2004ApJ_615_645}
{Onken}, C.~A., {Ferrarese}, L., {Merritt}, D., {et~al.} 2004, \apj, 615, 645

\bibitem[{{Poncelet} {et~al.}(2006){Poncelet}, {Perrin}, \&
  {Sol}}]{Poncelet2006AA.450.483}
{Poncelet}, A., {Perrin}, G., \& {Sol}, H. 2006, \aap, 450, 483

\bibitem[{{Przygodda} {et~al.}(2003){Przygodda}, {Chesneau}, {Graser},
  {Leinert}, \& {Morel}}]{Przygodda2003ApSS_286_85}
{Przygodda}, F., {Chesneau}, O., {Graser}, U., {Leinert}, C., \& {Morel}, S.
  2003, \apss, 286, 85

\bibitem[{{Reeves} {et~al.}(2004){Reeves}, {Nandra}, {George}, {Pounds},
  {Turner}, \& {Yaqoob}}]{Reeves2004ApJ_602_648}
{Reeves}, J.~N., {Nandra}, K., {George}, I.~M., {et~al.} 2004, \apj, 602, 648

\bibitem[{{Reichert} {et~al.}(1994){Reichert}, {Rodriguez-Pascual}, {Alloin},
  {Clavel}, {Crenshaw}, {Kriss}, {Krolik}, {Malkan}, {Netzer}, {Peterson},
  {Wamsteker}, {Altamore}, {Altieri}, {Anderson}, {Blackwell}, {Boisson},
  {Brosch}, {Carone}, {Dietrich}, {England}, {Evans}, {Filippenko}, {Gaskell},
  {Goad}, {Gondhalekar}, {Horne}, {Kazanas}, {Kollatschny}, {Koratkar},
  {Korista}, {MacAlpine}, {Maoz}, {Mazeh}, {McCollum}, {Miller}, {Mendes de
  Oliveira}, {O'Brien}, {Pastoriza}, {Pelat}, {Perez}, {Perola}, {Pogge},
  {Ptak}, {Recondo-Gonzalez}, {Rodriguez-Espinosa}, {Rosenblatt}, {Sadun},
  {Santos-Lleo}, {Shields}, {Shrader}, {Shull}, {Simkin}, {Sitko}, {Snijders},
  {Sparke}, {Stirpe}, {Stoner}, {Storchi-Bergmann}, {Sun}, {Wang}, {Welsh},
  {White}, {Winge}, \& {Zheng}}]{Reichert1994ApJ_425_582}
{Reichert}, G.~A., {Rodriguez-Pascual}, P.~M., {Alloin}, D., {et~al.} 1994,
  \apj, 425, 582

\bibitem[{{Reunanen} {et~al.}(2003){Reunanen}, {Kotilainen}, \&
  {Prieto}}]{Reunanen2003MNRAS_343_192}
{Reunanen}, J., {Kotilainen}, J.~K., \& {Prieto}, M.~A. 2003, \mnras, 343, 192

\bibitem[{{Roche} {et~al.}(1991){Roche}, {Aitken}, {Smith}, \&
  {Ward}}]{Roche1991MNRAS_248_606}
{Roche}, P.~F., {Aitken}, D.~K., {Smith}, C.~H., \& {Ward}, M.~J. 1991, \mnras,
  248, 606

\bibitem[{{Schmitt} {et~al.}(2003){Schmitt}, {Donley}, {Antonucci},
  {Hutchings}, \& {Kinney}}]{Schmitt2003ApJS_148_327}
{Schmitt}, H.~R., {Donley}, J.~L., {Antonucci}, R.~R.~J., {Hutchings}, J.~B.,
  \& {Kinney}, A.~L. 2003, \apjs, 148, 327

\bibitem[{{Swain} {et~al.}(2003){Swain}, {Vasisht}, {Akeson}, {Monnier},
  {Millan-Gabet}, {Serabyn}, {Creech-Eakman}, {van Belle}, {Beletic},
  {Beichman}, {Boden}, {Booth}, {Colavita}, {Gathright}, {Hrynevych},
  {Koresko}, {Le Mignant}, {Ligon}, {Mennesson}, {Neyman}, {Sargent}, {Shao},
  {Thompson}, {Unwin}, \& {Wizinowich}}]{Swain2003ApJ.596L.163}
{Swain}, M., {Vasisht}, G., {Akeson}, R., {et~al.} 2003, \apjl, 596, L163

\bibitem[{{Theureau} {et~al.}(1998){Theureau}, {Bottinelli}, {Coudreau-Durand},
  {Gouguenheim}, {Hallet}, {Loulergue}, {Paturel}, \&
  {Teerikorpi}}]{Theureau1998AAS_130_333}
{Theureau}, G., {Bottinelli}, L., {Coudreau-Durand}, N., {et~al.} 1998, \aaps,
  130, 333

\bibitem[{{Tristram} {et~al.}(2007){Tristram}, {Meisenheimer}, {Jaffe},
  {Schartmann}, {Rix}, {Leinert}, {Morel}, {Wittkowski}, {R{\"o}ttgering},
  {Perrin}, {Lopez}, {Raban}, {Cotton}, {Graser}, {Paresce}, \&
  {Henning}}]{Tristram2007AA.474.837}
{Tristram}, K.~R.~W., {Meisenheimer}, K., {Jaffe}, W., {et~al.} 2007, \aap,
  474, 837

\bibitem[{{Weingartner} \& {Draine}(2001)}]{Weingartner2001ApJ_548_296}
{Weingartner}, J.~C. \& {Draine}, B.~T. 2001, \apj, 548, 296

\end{thebibliography}

\end{document}